\documentclass[11pt,twoside]{article} 
\usepackage{amsmath,amssymb,latexsym}

\newcommand{\M}[1]{\mathcal{#1}}
\newcommand{\ds}{\displaystyle} 
\newcommand{\fract}[2]{{\textstyle\frac{#1}{#2}}} 
\DeclareMathOperator{\diag}{diag}
\begin{document}
\pagestyle{myheadings}
\markboth{P.C.~Schuster and R.L. Jaffe}{Quantum Mechanics on Manifolds Embedded in
Euclidean Space}
\title{\Large\bfseries Quantum Mechanics on Manifolds\\ Embedded in Euclidean
Space}
\author{P.C.~Schuster\footnote{Email: schuster@mit.edu\quad   Homepage:
http://web.mit.edu/schuster/Public/homepage.html}\enspace
and
R.L.~Jaffe\footnote{Email: jaffe@lns.mit.edu}\\
                     \\
{\small\itshape Center for Theoretical Physics and Department of Physics}  \\[-1ex]
{\small\itshape Massachusetts Institute of Technology} \\[-1ex] 
{\small\itshape  Cambridge, Massachusetts 02139} }
\date{\small MIT-CTP-3344} 
\maketitle
\begin{abstract}
\noindent 
Quantum particles confined to surfaces in higher dimensional spaces are acted
upon by forces that exist only as a result of the surface geometry and the
quantum mechanical nature of the system. The dynamics are particularly rich
when confinement is implemented by forces that act normal to the surface. We
review this \emph{confining potential formalism} applied to the confinement
of a particle to an arbitrary manifold embedded in a higher dimensional
Euclidean space. We devote special attention to the geometrically induced
gauge potential that appears in the effective Hamiltonian for motion on the
surface. We emphasize that the gauge potential is only present when the space
of states describing the degrees of freedom normal to the surface is
degenerate. We also distinguish between the effects of the intrinsic and
extrinsic geometry on the effective Hamiltonian and provide simple
expressions for the induced scalar potential. We discuss examples including
the case of a 3-dimensional manifold embedded in a 5-dimensional Euclidean
space. 
\end{abstract}


\section{Introduction}

In quantum mechanics the problem of constraining particle motion to a spatial manifold embedded
in a Euclidean space $R^n$ is conventionally treated in one of two ways. In the {\it intrinsic
quantization} approach, the motion is constrained to the manifold a priori. A classical Hamiltonian is
constructed from coordinates and momentum intrinsic to the surface and the system is quantized
canonically. In this case, the embedding space $R^n$ is irrelevant and the quantum system depends
only on the geometry intrinsic to the manifold. In the {\it confining potential} approach, the particle
is confined by a strong force that acts normal to the manifold. An effective Hamiltonian for
propagation on the hypersurface is obtained by freezing the motion normal to the surface in a low
state of excitation of the confining potential. This effective Hamiltonian depends on the intrinsic
geometry \emph{and} on the way that the hypersurface is embedded in $R^n$. The intrinsic
quantization scheme suffers from ordering ambiguities that allow for multiple consistent quantization
procedures that differ by a term proportional to the scalar curvature of the
hypersurface~\cite{ref:1.21}. On the other hand, the confining potential approach yields a unique
effective Hamiltonian that depends on the physical mechanism of the constraint. In any real physical
system we know of, constrained motion is the result of a strong confining force, and so one can argue
that the confining potential formalism offers a physically more realistic model of constraints.
Although intrinsic quantization has been studied since the earliest days of quantum theory, the
confining potential approach has only received serious attention in the last decade or two
\cite{ref:1.22,ref:1.4,ref:1.3}. 

The confining potential approach has now been studied for a variety of
systems using a variety of different confining forces. It has been
applied to the study of both spinless and spin-$\frac{1}{2}$ particles
confined to thin tubes and especially under the assumption that the
curvature of the tube is small and slowly
varying~\cite{ref:1.3,ref:1.17}. Spinless particles confined to
surfaces in three dimensional space have also been studied by da~Costa~\cite{ref:1.4}, and
the generalization to an arbitrary $m$-dimensional manifold $M^m$ embedded in a
$n$-dimensional Euclidean space has been carried out by subsequent
authors~\cite{ref:1.6,ref:1.1,ref:1.2,ref:1.7}. Extensions of the
confining potential approach have been applied to solitons with
confined collective degrees of freedom~\cite{ref:1.7} and to systems
for which the effective Hamiltonian on the hypersurface admits
supersymmetric states~\cite{ref:1.5}. Notable applications of the
confining potential formalism include the study of rotational spectra
of molecules~\cite{ref:1.1} and the study of electrons in Quantum Hall
devices~\cite{ref:1.18}. In this paper, we review and develop the
confining potential formalism in the spirit of Refs.~\cite{ref:1.7} and
\cite{ref:1.6}. We devote special attention to the group-theoretic
structure of the torsion-dependent terms that appear in the effective
Hamiltonian, as well as the role played by symmetries of the confining
potential. We also recast the mathematical form of the curvature-dependent potentials found by
previous authors in terms of ``principal curvatures'' so that the effects of the embedding structure on
the effective dynamics can be more easily understood.

Our paper is organized as follows. In Section 2, we briefly discuss the
physical motivation behind the confining potential approach. In Section 3, we
introduce an adapted coordinate system that allows one to separate normal
from tangential degrees of freedom on $M^m$. In Section 4, we derive the
effective Hamiltonian governing motion on $M^m$ by rescaling the collective
normal coordinates and developing a perturbative expansion in small
parameters of the complete Hamiltonian as found in \cite{ref:1.1}. In Section
5, we discuss in detail the gauge structure of the effective theory with
special attention devoted to the representation content of the gauge fields.
In Section 6, we discuss the intrinsic versus extrinsic geometrical
contributions to the effective theory on $M^m$ and give examples that
illustrate the possible physics for $M^3$ embedded in $R^{n\geq 4}$. Section~7 concludes with final
thoughts and discussion.

\section{The Confining Potential Approach}
We limit ourselves to the case of a scalar confining potential, although
magnetic-like vector potentials and gravitational-like tensor potentials are
physically relevant in some instances and have received attention in the
literature~\cite{ref:1.13}. A strong confining potential is introduced in all
directions normal to the hypersurface. The effect of this potential is to
constrain the particle to the manifold by raising the energy of normal
excitations far beyond the energy scale associated with motion tangent to the
hypersurface. The Hamiltonian then separates into a term governing high
energy confined motion in the directions normal to $M^m$, a term governing
low energy motion tangent to $M^m$, and interaction terms that couple the
normal and the tangential degrees of freedom. To obtain an effective
Hamiltonian on $M^m$, the total Hamiltonian is projected onto a low-lying multiplet of normal states,
typically the ground state. The effective Hamiltonian governing dynamics on $M^m$ is found to be the
Laplacian on $M^m$ coupled minimally to a background gauge field plus a
scalar quantum effective potential that depends on the principal
curvatures of $M^m$\cite{ref:1.4}. The gauge group is whatever subgroup
of SO($n-m=p$) is preserved by the confining potential. The strength
and representation content of the gauge terms appearing in the
effective theory depend not only on the properties of the embedding of
$M^m$, but also crucially on the symmetries of the space of normal
states. When the normal space is trivial, the gauge interaction in the
effective Hamiltonian vanishes identically. Only in cases where the
normal space is nontrivial (i.e., possesses degeneracies) will the
gauge interaction be nonzero. Thus static external SO($p$) gauge fields
can be geometrically induced by confining particles to manifolds that
are embedded nontrivially in a higher dimensional Euclidean space
using confining potentials that admit a degenerate space of normal
states.

\section{Geometry}
To study the quantum mechanics of a spinless particle confined to an
$m$-dimensional manifold $M^m$ embedded in $n$-dimensional Euclidean space $R^n$,
we first define a coordinate system that facilitates the separation of the
degrees of freedom normal to $M^m$ from those that are tangential. With
$\mathbf{R}:M^m\longrightarrow R^n$ denoting an embedding of $M^m$ in $R^n$,
and {$x^{\mu}, \mu=1,\ldots ,m$} a local coordinate system on $M^m$, we introduce
an {\it adapted coordinate frame} $\cal{F}$ defined by a smooth assignment of
$m$ linearly independent tangent vectors
$\mathbf{t}_{\mu}=\partial_{\mu}\mathbf{R}$, and $n-m=p$ orthogonal normal
vectors $\mathbf{\hat{n}}^i(x), i=m+1,\ldots ,n$. In a sufficiently small
neighborhood of $M^m$, the Cartesian coordinates, $\mathbf{r}$, for a point
in $R^n$ can be reexpressed as
\begin{equation}
\mathbf{r}(x,y)=\mathbf{R}(x) + y^i\mathbf{\hat{n}}^i(x) 
\label{2.1}
\end{equation}
where $x$ denotes an appropriate set of the $x^{\mu}$ and $y$ a set of distances $y^i$ from $M^m$
in the directions $\mathbf{\hat{n}}^i(x)$. The metric in the frame $\cal{F}$ is defined by
\begin{equation}
G_{AB}\equiv \partial_A\mathbf{r}\cdot \partial_B\mathbf{r}
\label{2.1b}
\end{equation}
where $A,B=1,\ldots,n$, and derivatives are taken with respect to adapted frame coordinates,
$x^{\mu}$ and $y^i$. To calculate $G_{AB}$, we need expressions for
$\partial_{\mu}\mathbf{\hat{n}}^i$ and
$\partial_{\mu}\mathbf{t}_{\nu}$.  Applying a generalized form of the Frenet-Serret equations
\cite{ref:1.19}, we may write
\begin{align}
\partial_{\mu}\mathbf{\hat{n}}^i     &= -\alpha^{i
\nu}_{\mu}\mathbf{t}_{\nu} - A^{ij}_{\mu}\mathbf{\hat{n}}^j \nonumber
\\
\partial_{\mu}\mathbf{t}_{\nu} &=
\Gamma^{\rho}_{\mu\nu}\mathbf{t}_{\rho} +
\alpha^i_{\mu\nu}\mathbf{\hat{n}}^i
\label{2.1c}
\end{align} 
where
\begin{align}
 g_{\mu\nu}        &= \mathbf{t}_{\mu}\cdot \mathbf{t}_{\nu} 
\nonumber\\
 \alpha_{\mu\nu}^i &= \mathbf{t}_{\mu}\cdot
\partial_{\nu}\mathbf{\hat{n}}^i \nonumber\\
 A^{ij}_{\mu}      &= \mathbf{\hat{n}}^i\cdot
\partial_{\mu}\mathbf{\hat{n}}^j    
\label{2.2}
\end{align} and ``$\cdot$'' is the standard inner product on $R^n$. In
the language of differential geometry, $g_{\mu\nu}$,
$\alpha^i_{\mu\nu}$, and $A^{ij}_{\mu}$ are the first fundamental form
(the metric on $M^m$), the second fundamental form, and the normal
fundamental form, respectively. The $\Gamma^{\rho}_{\mu\nu}$ are the
usual Christoffel symbols, but they will play little role in the
discussion that follows. We follow conventional notation in that lower
indices on tensors are obtained from upper indices by contraction with
$g_{\mu\nu}$.

Note that the choice of an adapted coordinate frame is not unique. In
particular, one adapted coordinate frame is carried into another by a
point-dependent rotation of the $\mathbf{\hat n }^i$. Under the action of
a rotation, $R^{ij}(x)$, on the normal vectors $\mathbf{\hat{n}}^i$,
$\alpha_{\mu\nu}^i$ transforms as an $SO(p)$ vector, and $A^{ij}_{\mu}$
as an $SO(p)$ gauge connection
\begin{equation}
A^{ij}_{\mu} \longrightarrow R^{ik}A^{kl}_{\mu}R^{jl} + R^{ik}\partial_{\mu}R^{jk}. 
\label{2.3}
\end{equation}
The metric can be determined from eqs.~(\ref{2.1b}) and (\ref{2.1c}), and is given by
\begin{equation}
G_{AB}= \begin{pmatrix} \gamma_{\mu\nu} + y^ky^lA^{kh}_{\mu}A^{lh}_{\nu} & y^kA^{jk}_{\mu} \\ y^kA^{ik}_{\nu} & \delta^{ij} \end{pmatrix},
\label{2.4}
\end{equation}
where $\gamma_{\mu\nu}$ is given by
\begin{equation}
\gamma_{\mu\nu} = g_{\mu\nu} - 2y^k\alpha^k_{\mu\nu} + y^ky^l\alpha^k_{\mu\rho}g^{\rho\sigma}\alpha^l_{\sigma\nu} .
\label{2.5}
\end{equation}
Calculating the determinant of $G_{AB}$, we find $|G|=|\gamma|$, where $|\gamma|$ is the determinant of $\gamma_{\mu\nu}$. Moreover, a calculation of the inverse of the metric tensor yields the exact expression, 
\begin{equation}
G^{AB} = \begin{pmatrix} \lambda^{\mu\nu} & \lambda^{\mu\sigma}y^kA^{kj}_{\sigma} \\ \lambda^{\nu\sigma}y^kA^{ki}_{\sigma} & \delta^{ij} + y^ky^lA^{ik}_{\sigma}A^{jl}_{\rho}\lambda^{\sigma\rho} \end{pmatrix}
\label{2.6}
\end{equation}
where $\lambda^{\mu\nu}\equiv (\gamma^{-1})_{\mu\nu}$ is the inverse of
$\gamma_{\mu\nu}$.\footnote{It should be noted that in the literature
$\lambda^{\mu\nu}$ is sometimes confused with $\gamma^{\mu\nu}$.}

\section{Derivation of the Effective Hamiltonian}

Having developed a convenient characterization of the geometry of $M^m$
embedded in $R^n$, we can construct the Hamiltonian $H$ on $M^m$. The
quantum description of dynamics on $M^m$ is unambiguously defined by
the free Hamiltonian $H_E$ on $R^n$ together with the potential that
confines the particle to $M^m$. To obtain the effective Hamiltonian on
$M^m$, we rewrite $H_E$ in terms of our adapted frame coordinates, and
then project it onto the space of states describing the confined normal
degrees of freedom. In Cartesian coordinates {$\mathbf{r}^A$}, and
working in units where $\hbar$ and the mass of our particle is equal to
unity, we have   
\begin{equation}
H_E = -\fract{1}{2}\partial^E_A\partial^E_A + V
\label{2.7}
\end{equation}
where $\partial^E$ denotes derivatives taken with respect to Euclidean coordinates, and $V\equiv V(y)$ is the confining potential that depends only on the normal coordinates.
Following convention, we normalize the wave function $\Phi$ of the system according to the condition
\begin{equation}
\int{ |\Phi|^2\,d^nr } = 1 .
\label{2.8}
\end{equation}
Changing coordinates to $x$ and $y$, the Hamiltonian given in eq.~(\ref{2.7}) becomes
\begin{equation}
H_E = -\frac{1}{2|G|^{1/2}}\partial_AG^{AB}|G|^{1/2}\partial_B + V
\label{2.9}
\end{equation}
and the normalization condition of eq.~(\ref{2.8}) becomes
\begin{equation}
\int{ |\Phi|^2|G|^{1/2}d^mx\,d^py } = 1 .
\label{2.10}
\end{equation}
Since we want to obtain a wave function describing a quantum mechanical probability density
for a particle moving on $M^m$, we rescale the wave function $\Phi$ by
$ |G|^{1/4} /\,|g|^{1/4}$, where $|g|$ is the determinant of $g_{\mu\nu}$, $\Psi\equiv
 (|G|^{1/4}/\,|g|^{1/4})\Phi$. Likewise, we rescale the Hamiltonian 
$H_E$, $H\equiv (|G|^{1/4}/\,|g|^{1/4}) H_E(|g|^{1/4}/\,|G|^{1/4})$. $\Psi$ is then normalized
on
$M^m$ as 
\begin{equation}
\int{d^m x\,|g|^{1/2}d^py\,|\Psi|^2 } = 1
\label{2.11b}
\end{equation}
and so $\int d^py\,|\Psi|^2$ can be interpreted as a probability density for a particle moving
on $M^m$ defined with respect to the conventional manifold measure $d^mx|\,g|^{1/2}$.
Returning to $H$, we may use the explicit form for $G^{AB}$ in eq.~(\ref{2.6}) to obtain
\begin{align}
H &=
-\frac{1}{2|\gamma|^{1/4}}\partial_i|\gamma|^{1/2}\partial_i\frac{1}{|\gamma|^{1/4}}\nonumber
\\ &\qquad
{}-\frac{1}{2|g|^{1/4}|\gamma|^{1/4}}
\biggl(\partial_{\mu}\lambda^{\mu\nu}|\gamma|^{1/2}\partial_{\nu}
+ y^ky^lA^{ik}_{\mu}A^{jl}_{\nu}\partial_{i}\lambda^{\mu\nu}|\gamma|^{1/2}\partial_{j} 
\label{2.12} \\
    &\qquad\qquad
{}+  \partial_{\mu}\lambda^{\mu\rho}y^kA^{kj}_{\rho}|\gamma|^{1/2}\partial_{j} +
\partial_{i}\lambda^{\nu\rho}y^kA^{ki}_{\rho}|\gamma|^{1/2}\partial_{\nu}\biggr)\frac{|g|^{1/4}}{|\gamma|^{1/4}}
+ V(y) . 
\nonumber
\end{align}
Introducing $\hat{\partial}_{\mu} \equiv \partial_{\mu} + \frac{1}{2}iA^{ij}_{\mu}L_{ij}$,
where the $L_{ij} = i(y^j\partial_i - y^i\partial_j)$ are the angular momentum operators  in the
space normal to $M^m$, we may compactly rewrite eq.~(\ref{2.12}) as
\begin{equation}
H = -\frac{1}{2|\gamma|^{1/4}}\partial_i|\gamma|^{1/2}\partial_i\frac{1}{|\gamma|^{1/4}} - \frac{1}{2|g|^{1/4}|\gamma|^{1/4}}\hat{\partial}_{\mu}\lambda^{\mu\nu}|\gamma|^{1/2}\hat{\partial}_{\nu}\frac{|g|^{1/4}}{|\gamma|^{1/4}} + V(y) .
\label{2.13}
\end{equation}

Next we implement the constraint imposed by the confining potential $V$. To do this, we
exploit the fact that $V$ is a function of the normal coordinates $y^i$ alone and $V$ has a deep
minimum at $y^i=0$. Thus, we may expand $V$ as a power series in the $y^i$ about its
minimum,  
\begin{equation}
V(y^i) = \fract{1}{2}\omega^2y^{i2} + O(y^3)
\label{2.14}
\end{equation}
where we have assumed that $V$ is symmetric in the $y^i$ up to quadratic
order.\footnote{Asymmetric scalar confining potentials are considered in Ref. \cite{ref:1.6}}
Since $V$ has a deep minimum, we can neglect terms of order $y^3$ and higher. In
neglecting these terms, we are assuming that $\omega$ is much larger than the scale of
curvatures on $M^m$, denoted by $\kappa$. More specifically, $\omega\gg\kappa^2$.
Following the approach of Refs.~\cite{ref:1.1} and \cite{ref:1.2}, we adsorb the scale of the
frequency $\omega$ in eq.~(\ref{2.14}) into a small dimensionless parameter $\epsilon$,
$\omega\rightarrow  \omega/\epsilon$, so that the rescaled $\omega$ is of order
$\kappa^2$. We then use $\epsilon$ as a natural perturbative parameter in the theory. Thus,
the dominant pieces of the Hamiltonian in eq.~(\ref{2.13}) that act on the transverse space
are
\begin{equation}
H_0 = -\frac{1}{2}\partial_i\partial_i + \frac{1}{2\epsilon^2}\omega^2y^{i2}.
\label{2.14a}
\end{equation}
Formally, we want to consider the limit $\epsilon\rightarrow 0$. However, the divergence in
the potential $(1/2\epsilon^2)\omega^{i2}y^{i2}$ in the $\epsilon\rightarrow 0$ limit
complicates the analysis. To avoid this problem, we rescale the coordinates $y^i$, as
$y^i\rightarrow \epsilon^{1/2}y^i$, which allows us to rewrite eq.~(\ref{2.14a}) as
\begin{equation}
H_0 = \frac{1}{\epsilon}(-\frac{1}{2}\partial_i\partial_i + \frac{1}{2}\omega^2y^{i2}) .
\label{2.14c}
\end{equation}
Thus, we can study the $\epsilon\rightarrow 0$ limit unambiguously by considering $\epsilon H$. 
We apply this approach to the complete Hamiltonian to develop an expansion of $\epsilon H$ in
powers of $\epsilon$, 
\begin{equation}
\epsilon H = \hat{H}_0 + \epsilon\hat{H} + O(\epsilon^{3/2})
\label{2.14d}
\end{equation}
where
\begin{equation}
\hat{H}_0 = \fract{1}{2}\left(-\partial_i\partial_i + \omega^{i2}y^{i2} \right)
\label{2.15}
\end{equation}
and\footnote{To obtain eq.~(\ref{2.16}) for $\hat{H}$, we have used
$\lambda^{\mu\nu}=g^{\mu\nu}+2\epsilon^{1/2}y^k\alpha^{k\mu\nu}+3\epsilon
y^ky^l\alpha^{l\rho\nu}\alpha^{k\mu}_{\rho}+O(\epsilon^{3/2})$.}
\begin{equation}
\hat{H} = -\frac{1}{2g^{1/2}}\Bigl(\partial_{\mu} + \frac{i}{2}A^{ij}_{\mu}L_{ij}
\Bigr)g^{\mu\nu}g^{1/2}\Bigl(\partial_{\nu} + \frac{i}{2}A^{kl}_{\nu}L_{kl} \Bigr) +
\fract{1}{8}g^{\mu\nu}g^{\rho\sigma}\left( \alpha^i_{\mu\nu}\alpha^i_{\rho\sigma} -
2\alpha^i_{\mu\rho}\alpha^i_{\nu\sigma} \right) .
\label{2.16}
\end{equation}
Equation (\ref{2.16}), which forms the basis of the subsequent analysis, was first obtained in full generality by Maraner and Destri~\cite{ref:1.6}. Given that we are interested in the $\epsilon\rightarrow 0$ limit, the only term beyond $\hat{H}_0$ relevant in the
perturbative expansion in eq.~(\ref{2.14d}) is $\hat{H}$. From here on we keep only $\hat{H}$ which
survives as
$\epsilon\rightarrow 0$.

To obtain an effective Hamiltonian on $M^m$, we need to ``freeze'' the normal degrees of freedom.
We separate the wave function $\Psi$ into a function depending on the normal coordinates $y^i$, and
a function depending on the manifold coordinates $x^{\mu}$
\begin{equation}
\Psi(x,y) = \sum_{\beta}\psi_{\beta}(x) \chi_{\beta}(y)
\label{2.19a}
\end{equation}
where the index $\beta=1,\ldots,d$ labels any degeneracy that exists in the spectrum of the
$O({1}/{\epsilon})$ Hamiltonian ${\hat{H}_0}/{\epsilon}$ governing the normal degrees of
freedom. $\hat{H}_0$ is degenerate because of the SO($p$) symmetry of $V(y)$, and so the
eigenstates of $\hat{H}_0$ can be decomposed into irreducible SO($p$) multiplets. For the case of a
$p$-dimensional symmetric harmonic oscillator, the ground state of $\hat{H}_0$ belongs to the trivial
representation of SO($p$), while the first excited state belongs to the $p$-dimensional ``fundamental''
representation of SO($p$). The $\chi_{\beta}(y)$ satisfy to $O(1/\epsilon)$
\begin{equation}
\frac{1}{\epsilon}\hat{H}_0\chi_{\beta}(y) = E_0\chi_{\beta}(y)
\label{2.19b}
\end{equation}
where $E_0$ gives the largest $O(1/\epsilon)$ contribution to the total energy $E$ of the
system. Upon projection onto the space of states spanned by $\chi_1(y),\ldots ,\chi_d(y)$, $\hat{H}$
becomes a $d\times d$ matrix $\M{\hat{H}}$ with components
\begin{equation}
\M{\hat{H}}_{\alpha\beta} = \int d^py\, \chi_{\alpha}^*(y)\hat{H}\chi_{\beta}(y) .
\label{2.20A}
\end{equation}
$\M{\hat{H}}$ acts on the wave function $\vec{\psi}(x)$ (with components $\psi_{\beta}(x)$), and the dynamics on $M^m$
is determined by
\begin{equation}
\M{\hat{H}}\vec{\psi}(x) = \hat{E}\vec{\psi}(x)
\label{2.20B}
\end{equation}
where $\hat{E}$ is the $O(\epsilon^0)$ correction to the total energy $E$ of the system.

\section{Gauge Structure}

To better understand the structure of the effective Hamiltonian $\M{\hat{H}}$, we return to expression (\ref{2.16}). 
Defining $d\times d$ matrices $\M{L}_{ij}$ and $\M{L}^2_{ij,kl}$ by
\begin{align}
(\M{L}_{ij})_{mn}      &\equiv \int d^py\, \chi^*_m(y)L_{ij}\chi_n(y)\nonumber \\
(\M{L}^2_{ij,kl})_{mn} &\equiv \int d^py\, \chi^*_m(y)L_{ij}L_{kl}\chi_n(y)
\label{2.20c}
\end{align}
and using eq.~(\ref{2.20A}), the effective Hamiltonian on $M^m$ can be rewritten as
\begin{equation}
\M{\hat{H}} = -\frac{1}{2g^{1/2}}(\mathbf{\partial}_{\mu} - i\M{A}_{\mu})g^{\mu\nu}g^{1/2}(\mathbf{\partial}_{\nu} - i\M{A}_{\nu}) + \M{P}
\label{2.21}
\end{equation}
where $\M{P}$ and $\M{A}_{\mu}$ are the $d\times d$ matrices
\begin{align}
\M{P}       &= \fract{1}{8}g^{\mu\nu}g^{\rho\sigma}\left(
\alpha^i_{\mu\nu}\alpha^i_{\rho\sigma} - 2\alpha^i_{\mu\rho}\alpha^i_{\nu\sigma}
\right)\M{I}  \nonumber \\
\M{A}_{\mu} &= \fract{1}{2}A^{rs}_{\mu}\M{L}_{sr}  
\label{2.22}
\end{align}
and $\M{I}$ is the $d\times d$ identity matrix. 

The algebra leading to eq.~(\ref{2.21}) generates a term $\M{W} = \frac{1}{8}g^{\mu\nu}A^{ij}_{\mu}A^{kl}_{\nu}(\M{L}^2_{ij,kl} - \M{L}_{ij}\M{L}_{kl})$ which is kept explicitly by other workers. Since the confining potential $V(y)$ possesses a SO($p$) symmetry, all of the $\M{L}_{ij}$'s commute with $\M{\hat{H}}$ and $\{\chi_{\beta}(y)\}$ forms a complete set of
states for the subspace spanned by $L_{ij}\chi_1(y),\ldots, L_{ij}\chi_{d}(y)$ for all $i,j$.
Consequently,
$\M{W}$ is zero.
       
$\M{\hat{H}}$ is the Hamiltonian for a spinless particle in a curved space in the presence of
background SO($p$) gauge fields and a geometrically induced potential. We emphasize that the gauge
potentials are only present if the normal wavefunction lies in a degenerate, nontrivial representation
of SO($p$). The effective physics on $M^m$ governed by $\M{\hat{H}}$ remains invariant under local
SO($p$) gauge transformations of the normal coordinates. Under SO($p$) rotations of the
$\mathbf{\hat{n}}^i$, $\M{A}_{\mu}$ transforms as a gauge field in the adjoint representation of
SO($p$), while $\vec{\psi}(x)$ transforms in some $d$-dimensional representation $D_d$ of SO($p$).
In particular, under the transformation
\begin{equation}
\mathbf{\hat{n}}^i\longrightarrow (\M{R})_{ij}\mathbf{\hat{n}}^j
\label{2.22b}
\end{equation}
where $\M{R}=e^{i\theta_{ij}\tilde{\M{L}}_{ij}}$ is an element of the vector representation of SO($p$), $\vec{\psi}(x)$, $\M{A}_{\mu}$, and $\M{P}$ transform as
\begin{align}
\vec{\psi}(x)&\longrightarrow \M{V}\vec{\psi}(x)\nonumber\\
\M{A}_{\mu}  &\longrightarrow \M{V}\M{A}_{\mu}\M{V}^T + \M{V}\partial_{\mu}\M{V}^T\nonumber \\
\M{P}        &\longrightarrow \M{P}
\label{2.22c}
\end{align}
where $\M{V}=e^{i\theta_{ij}\M{L}_{ij}}$ is in the $D_d$ matrix representation of SO($p$). As promised, the invariance of $V$ under SO($p$)-rotations is realized as an SO($p$) gauge invariance of the effective theory on $M^m$.

The field strength tensor, $\M{G}_{\mu\nu}$, associated with the gauge potential $\M{A}_{\mu}$ is given by
\begin{equation}
\M{G}_{\mu\nu} = \partial_{\mu}\M{A}_{\nu}-\partial_{\nu}\M{A}_{\mu}+[\M{A}_{\mu},\M{A}_{\nu}] .
\label{2.22d}
\end{equation}
Although a nonvanishing field strength is a sufficient condition for the gauge potential $\M{A}_{\mu}$ to have a physical effect, it is not necessary. As Takagi and Tanzawa have noted, even in cases with vanishing field strength, global Aharonov-Bohm effects can exist when
the constraint hypersurface has nonvanishing torsion \cite{ref:1.3}. Subsequent authors have further
explored the connection between Aharonov-Bohm effects and the geometry of the constraint
hypersurface for the case of $M^1$ embedded in $R^n$ \cite{ref:1.7}.

\section{Intrinsic and Extrinsic Geometry Contributions}
In this section, we consider two questions: First, what features of $\M{\hat{H}}$ \emph{cannot} be purely attributed to the intrinsic geometry of $M^m$? Second, what type of embedding structure is needed
to generate nontrivial geometrically induced physics effects for the case of $M^3$ embedded in
$R^{n\geq4}$? 

Upon examining eqs.~(\ref{2.21}) and (\ref{2.22}), we see that the \emph{intrinsic} contributions to
$\M{\hat{H}}$ are from the Laplacian on $M^m$ involving the adapted frame metric $g_{\mu\nu}$.
\emph{Extrinsic} contributions to $\M{\hat{H}}$ occur through the momentum independent potential
$\M{P}$ and through the minimally coupled gauge field $\M{A}_{\mu}$. $\M{P}$ depends on the
extrinsic geometry of the embedding of $M^m$ in $R^n$ and is purely quantum mechanical (i.e., does
not survive in the classical limit). $\M{P}$ was generated by rescaling the Hamiltonian to adaptive
coordinates, so it represents a quantum ``fictitious'' force associated with the adapted frame
$\cal{F}$. In order to understand better how the embedding generates the effective potential, we
rewrite $\M{P}$ in terms of the geometrically invariant principal curvatures of $M^m$. There are
$m$ principal curvatures for each normal vector $\mathbf{\hat{n}}^i$ given by the eigenvalues of the
matrix
\begin{equation}
(\hat{\alpha}_i)_{\mu\nu} = \alpha^{i \nu}_{\mu} .
\label{2.25b}
\end{equation}
Denoting the $\mu^{th}$ principal curvature corresponding to the $i^{th}$ normal $\mathbf{\hat{n}}^i$ as $\kappa_{\mu, i}$, we introduce the linear and quadratic polynomials symmetric in the $\kappa_{\mu, i}$ for each $i$ independently, 
\begin{equation}
s_{1,i} = \sum_{\mu} \kappa_{\mu,i}\hspace{1in} s_{2,i} = \sum_{\mu<\nu} \kappa_{\mu,i}\kappa_{\nu,i}   .
\label{2.25c}
\end{equation}
In terms of the $s_{1,i}$ and $s_{2,i}$, we have
\begin{align}
\M{P} &= \fract{1}{8}\sum_i\left( \mathtt{tr}(\hat{\alpha}_i)^2 - 2\mathtt{tr}(\hat{\alpha}_i^2)
\right)\M{I}\nonumber \\
      &= -\fract{1}{8}\sum_i\left( s_{1,i}^2 - 4s_{2,i} \right)\M{I} .
\label{2.18}
\end{align}
As first pointed out in Refs.~\cite{ref:1.3} and \cite{ref:1.4}, the effective potential for the
cases of $M^{1,2}$ embedded in $R^3$ is given by, 
\begin{align}
\M{P} &= -\fract{1}{8}\kappa^2\M{I} \hspace{.5in} (M^1)\nonumber, \\
\M{P} &= -\fract{1}{8}(\kappa_1 - \kappa_2)^2\M{I} \hspace{.5in} (M^2), 
\label{2.26}
\end{align}
where $\kappa$ is the curvature of the curve $M^1$, and $\kappa_1, \kappa_2$ are the principal curvatures of $M^2$. For both of these cases, the effective potential is strictly negative. For the more general case of $M^{m\geq 3}$ embedded in $R^{n\geq m+1}$, the effective potential can locally equal any real valued smooth function defined on $M^m$. 
For example, with $\kappa_{1,2,3}$ denoting principle curvatures
\begin{equation}
\M{P} = -\fract{1}{8}\Bigl(\kappa_3\bigl(\kappa_3-2(\kappa_1+\kappa_2)\bigr) +
(\kappa_1-\kappa_2)^2\Bigr)\M{I} 
\label{2.27}
\end{equation}
for the case of $M^3$ embedded in $R^4$. 

The ``fictitious force'' potential $\M{P}$ is nonzero for a general embedding of $M^3$ in $R^n$. The
other geometry-dependent interaction that can appear in $\M{\hat{H}}$ is the gauge term
$\M{A}_{\mu}$. For the simplest 3-dimensional case of $M^3$ embedded in $R^4$, $V$ is a
1-dimensional potential with nondegenerate energy eigenstates, and so $\M{A}_{\mu}$ vanishes.
For $M^3$ embedded in $R^{n\geq 5}$, $\M{A}_{\mu}$ can be nonzero when the $\chi_{\beta}(y)$
are locked into a degenerate subspace of $V$. In the degenerate cases, the gauge interaction has
U($1$) symmetry for $m=3$ and $n=5$, and SO($3$) symmetry for $m=3$ and $n=6$. 

To further illustrate the physics for a 3-dimensional manifold, consider the embedding of a
hypersurface $M^3$ in $R^5$ given by
\begin{equation}
\mathbf{R}(x,y,z)=(x\cos{\rho z},x\sin{\rho z},x,y,z) . 
\label{2.28}
\end{equation}
In this example, the $x$ and $z$ coordinates have been mapped onto a helical surface in a 3-dimensional subspace of $R^5$ and the remaining $y$ coordinate mapping is flat. Using an adapted frame field
\begin{align}
\mathbf{t}_x &= (\cos{\rho z},\sin{\rho z},1,0,0)\nonumber\\
\mathbf{t}_y &= (0,0,0,1,0)\nonumber\\
\mathbf{t}_z &= (-\rho x\sin{\rho z},\rho x\cos{\rho z},0,0,1)\nonumber\\
\mathbf{\hat{n}}^1 &= \frac{1}{\sqrt{2}}(\cos{\rho z},\sin{\rho z},-1,0,0)\nonumber\\
\mathbf{\hat{n}}^2 &= \frac{1}{\sqrt{1+\rho^2x^2}}(\sin{\rho z},-\cos{\rho z},0,0,\rho x)
\label{2.29}
\end{align}
we may calculate the nonvanishing components of the fundamental forms and obtain
$g_{\mu\nu} = \diag (1,1,1+\rho^2x^2)$, $\alpha_{\mu\nu}^1 =
\diag(0,0, {\rho^2x}/{\sqrt{2}})$, 
\begin{equation}
\alpha_{\mu\nu}^2 = \begin{pmatrix}  0 & 0 &
\ds \frac{\rho}{\sqrt{1+\rho^2x^2}}
\\ 0 & 0 & 0 \\ \ds \frac{\rho}{\sqrt{1+\rho^2x^2}} & 0 & 0
\end{pmatrix}
\label{2.29a}
\end{equation}
and $A^{12}_z  =   \rho / \sqrt{2+2\rho^2x^2}$.
Moreover, the field strength tensor $\M{G}_{\mu\nu}$ has nonvanishing components
\begin{equation}
\M{G}_{\mu\nu}=\begin{pmatrix} 0 & 0 &\ds -\frac{l\rho^3x}{\sqrt{2}(1+\rho^2x^2)^{3/2}} \\ 0 & 0
& 0 \\\ds \frac{l\rho^3x}{\sqrt{2}(1+\rho^2x^2)^{3/2}} & 0 & 0 \end{pmatrix}
\label{2.30}
\end{equation}
where $l$ is the expectation value of the normal state angular momentum operator
$\M{L}_{12}$. The induced potential $\M{P}$ can be obtained from
$\hat{\alpha}_1 = \diag(0,0,{\rho^2x}/{\sqrt{2}(1+\rho^2x^2)})$, 
\begin{equation}
\hat{\alpha}_2 =\begin{pmatrix} 0 & 0 &\ds \frac{\rho}{(1+\rho^2x^2)\sqrt{1+\rho^2x^2}} \\ 0
& 0 & 0 \\\ds \frac{\rho}{\sqrt{1+\rho^2x^2}} & 0 & 0 \end{pmatrix}
\label{2.30B}
\end{equation}
and is given by
\begin{equation}
\M{P} = -\frac{\rho^2}{8(1+\rho^2x^2)^2}\left(4+\fract{1}{2}\rho^2x^2 \right) \M{I}.
\label{2.31}
\end{equation}
Given a confining potential with Abelian SO($2$) $\cong$ U($1$) invariance, a nonvanishing field strength tensor implies that the U($1$) induced gauge potential $\M{A}_{\mu}$ cannot be transformed away. Assuming that the normal states are locked into a subspace with nonvanishing angular momentum $l$, the gauge potential is given by
\begin{equation}
\M{A}_{\mu} = \Bigl(0,0,\frac{\rho l}{\sqrt{2+2\rho^2x^2}}\Bigr)
\label{2.32}
\end{equation}
and the corresponding background magnetic-like field is
\begin{equation}
\M{B}_{\mu} = \nabla\times\M{A}_{\mu} =\Bigl (0,\frac{\rho^3x
l}{\sqrt{2}(1+\rho^2x^2)^{3/2}},0\Bigr) .
\label{2.33}
\end{equation} 
For fixed $x$, an observer on $M^3$ would feel the presence of a magnetic field along the
$y$-direction and an attractive $x$-dependent scalar potential centered at $x=0$, both of which tend
to zero as $x$ goes to infinity. 

In addition to curves and surfaces embedded in $R^3$, other examples that have received attention in the literature include SO($3$) embedded in $R^{3n}$ \cite{ref:1.1,ref:1.2}, as well as generalized curves $M^1$ and S$^m$ embedded in $R^n$ \cite{ref:1.7}. 

\section{Conclusion}

Confinement of particle motion to a curved manifold generates gauge fields as well as fictitious forces
in the effective theory on the manifold. We have applied the confining potential formalism to the
study of systems with confined degrees of freedom, and demonstrated that in the adiabatic limit of
slowly varying curvature, the strength and representation content of the gauge terms appearing in
the effective theory depends crucially on the space of normal states. The gauge terms vanish when
the normal state is nondegenerate.

In addition to gauge terms, fictitious forces that depend on the extrinsic geometry of the constraint
manifold also appear in the effective theory. The extrinsic geometric contributions to the theory
highlight a fundamental difference between confinement in classical versus quantum physics. In
classical mechanics, dynamics on the constraint manifold is independent of the directions normal to
the manifold and therefore depends only on intrinsic geometry. In quantum mechanics, the
Schr\"odinger wave function of the system is always nonzero in some neighborhood of the
constraint manifold and is therefore sensitive to both intrinsic and extrinsic geometry.

\vspace{.2cm}
\subsection*{Acknowledgments} 
This work is supported in part by funds provided by the U.S. Department of Energy (D.O.E.) under cooperative research agreement \#DF-FC02-94ER40818.

\vspace{.2cm}

\end{document}